\documentclass[12pt,cits,hyper]{JINST}

\usepackage{booktabs}
\usepackage{setspace}

\newcommand{\alp}{\ensuremath{\alpha}}

\newcommand{\mgcs}{\ensuremath{\,\mathrm{mg/cm^{2}}}}

\newcommand{\mus}{\ensuremath{\,\mu\mathrm{s}}}

\newcommand{\degr}{\ensuremath{^{\circ}}}

\title{Development of wavelength shifter
coated reflectors for the ArDM argon dark matter detector}
\author{V. Boccone$^a$, P.~K.~Lightfoot$^e$, K.~Mavrokoridis$^e$, C.~Regenfus$^a$, C.~Amsler$^a$, A.~Badertscher$^b$, A.~Bueno$^c$, H.~Cabrera$^a$, M.~C.~Carmona-Benitez$^c$,
M.~Daniel$^d$, E.~J.~Daw$^e$, U.~Degunda$^b$, A.~Dell'~Antone$^a$, A.~Gendotti$^b$, L.~Epprecht$^b$, 
S.~Horikawa$^b$, L.~Kaufmann$^b$,
L.~Knecht$^b$, M.~Laffranchi$^b$, C.~Lazzaro$^b$, D.~Lussi$^b$, J.~Lozano$^c$, A.~Marchionni$^b$,
A.~Melgarejo$^c$, P.~Mijakowski$^f$,
G.~Natterer$^b$, S.~Navas-Concha$^c$, P.~Otyugova$^a$, M.~de~Prado$^d$, P.~Przewlocki$^f$, 
F.~Resnati$^b$, M. Robinson$^e$, J.~Rochet$^a$, L.~Romero$^d$, E.~Rondio$^f$, 
A.~Rubbia$^b$ , N.~J.~C.~Spooner$^e$, T.~Strauss$^b$, J.~Ulbricht$^b$, T.~Viant$^b$ (The~ArDM~Collaboration)\\
\llap{$^a$}University of Z\"urich,
Physik-Institut, CH--8057 Z\"urich, Switzerland\\
\llap{$^b$}ETH Zurich,
Institute for Particle Physics, CH--8093 Z\"urich, Switzerland\\
\llap{$^c$}University of Granada,
Dpto. de F\'isica Te\'orica y del Cosmos \& C.A.F.P.E,
Campus Fuente Nueva, 18071 Granada, Spain\\
\llap{$^d$}CIEMAT, Div. de Fisica de Particulas,
Avda. Complutense, 22, E-28040, Madrid, Spain\\
\llap{$^e$}University of Sheffield,
Department of Physics and Astronomy, Hicks Building, Hounsfield Road, Sheffield, S3~7RH, UK\\
\llap{$^f$}The Andrzej Soltan Institute for Nuclear Studies, H\.oza 69, 00-681 Warsaw, Poland
}

\abstract{
To optimise the design of the light readout in the ArDM 1-ton liquid argon dark matter detector, a range
of reflector and WLS coating combinations were investigated in several small setups, where 
argon scintillation light was generated by radioactive sources in gas at normal temperature and 
pressure and shifted into the blue region by tetraphenyl butadiene (TPB). 
Various thicknesses of TPB were deposited by spraying and vacuum evaporation onto specular
3M{\small\texttrademark}-foil and diffuse Tetratex{\small\textregistered} (TTX) substrates. 
Light yields of each reflector and TPB coating combination were compared. 
Reflection coefficients of TPB coated reflectors were independently
measured using a spectroradiometer in a wavelength range between
200 and 650~nm.
WLS coating on the PMT window was also studied. 
These measurements were used to define the parameters of the light reflectors
of the ArDM experiment. Fifteen large $120\times 25$~cm$^2$ TTX sheets
were coated and assembled in the detector. Measurements in argon 
gas are reported providing good evidence of fulfilling the light collection
requirements of the experiment.
}


\begin{document}

\section{Introduction}

The ArDM experiment ~\cite{Rubbia:2005ge,Laffranchi:2007p1321}
aims for the operation of a ton-scale liquid argon 
target for direct detection of dark matter 
particles scattering off nuclei with a recoil energy threshold
of 30~keV. A high reconstruction sensitivity for such events depends on an efficient 
detection of scintillation light and ionization charge. Ionizing radiation in liquid noble gases leads to 
the formation of excimers in either singlet or triplet states~\cite{Doke:1990p1608,Kubota:1978p975}, which decay radiatively
to the dissociative ground state with characteristic fast and slow lifetimes  ($\tau_{fast}\approx 6$~ns, $\tau_{slow}\approx 1.6\mu$s in liquid argon
 with the so-called second continua emission spectrum peaked at 
$128\pm10$~nm~(\cite{larlight1}).
In warm argon gas at 1~bar $\tau_{slow}\approx 3.2\mu$s and
the third continuum in the wavelength range between 175 and 250~nm
populates most of the fast component~\cite{Langhoff:1988p1087,Krotz:1991p1082,Amsler:2008p1320}.
Singlet and triplet states are produced with different amplitudes 
depending on the ionizing radiation. In addition, the phenomenon of recombination
effectively transforming ionization into scintillation~\cite{Kubota:1978p964}, depends on the ionization
density in the medium.
Based on these properties, it was shown that the relative fraction of scintillation to
ionization~\cite{Benetti:1993} and  the time structure of the argon scintillation light~\cite{Boulay:2006mb} can be used in liquid
to discriminate nuclear recoils against $\gamma$ and electron backgrounds which enter
the sensitive fiducial region of the target.

\begin{figure}[tb]
\centering
a)\includegraphics[width=0.4\textwidth]{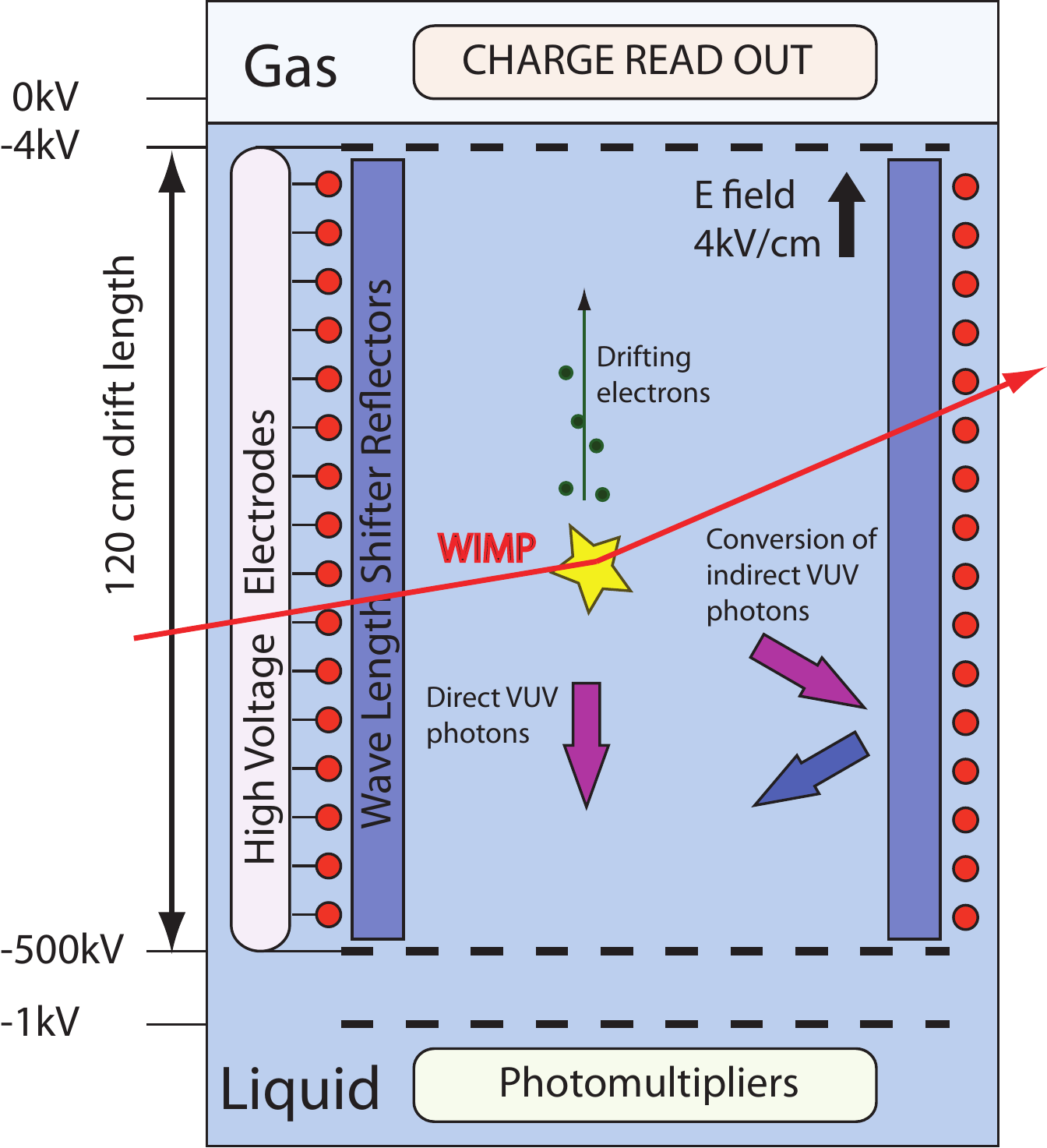}\hspace{18mm}b)\includegraphics[width=0.4\textwidth]{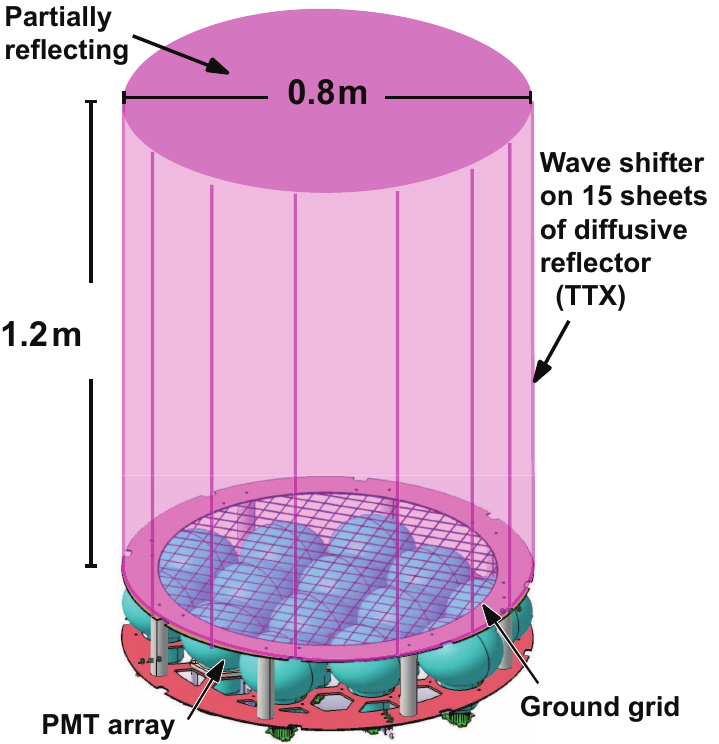}
\caption{\sl (left) Operation principle of the ArDM detector, (right) 3D sketch of the  83\,cm by 120\,cm cylindrical light detection volume.}
\label{fig:intro}
\end{figure}

The ArDM detector will be operated as a double phase LAr LEM-TPC~\cite{Badertscher:2008rf}, as is illustrated in Figure\,\ref{fig:intro}(left). 
The charge readout will be situated in the gas phase at the top of the vessel, while photodetectors sensitive to single photons will be located 
in the liquid argon at the bottom of the apparatus. In the following, we concentrate on the light readout system.

As mentioned above, the UV scintillation light of liquid
argon is below 150\,nm.
Since large VUV sensitive PMTs (e.g.\,MgF$_2$ windowed) 
are not commercially available, the use of reflectors coated with a wavelength shifter (WLS) 
along with standard bialkali photomultiplier tubes (PMTs) is an
economical realisation of an efficient readout system. 

In ArDM, the light readout system will be  composed of 14 hemispheric  8'' Hamamatsu R5912-02MOD-LRI
PMTs, made from particularly radiopure 
borosilicate glass and feature bialkali photocathodes with Pt-underlay 
for operation at cryogenic temperatures~\cite{bueno2008}. 
The argon scintillation light  is 
wavelength shifted into the range of maximum quantum efficiency (QE) of the PMT
by a thin layer~(1.0~$\mgcs$) of tetra\-phenyl\-buta\-diene 
(TPB) evaporated onto 15 cylindrically arranged reflector sheets 
(120$\times$25\,cm$^2$) which are located in the vertical electric field. 
These sheets are made of the PTFE fabric Tetratex{\small\textregistered} 
(TTX), transporting the shifted light diffusively onto the PMTs 
(shown in Figure\,\ref{fig:intro} (right)). 
The PMT glass windows are also coated with TPB to convert directly 
impinging DUV photons. 

%%--------------
%%----------------

In this paper, we present our investigations on TPB coating methods, sample preparation,
and optimisation of TPB coating thickness on two separate reflector
substrates. These results have led to the optimization
of the reflector and PMT coating used for the ArDM experiment.

\section{WLS,  reflector material and coating techniques}
\label{chap:expmeth}

\subsection{The wavelength shifting chemical}

We seek a fast response of the light readout system not to 
distort the pulse shape of the li\-quid argon scintillation light. 
Most organic wave shifting materials are well known 
for their fast optical response caused by the rapid
process of radiative recombination of electron hole pairs at 
the benzene rings in their chemical structure.
Typical well-known elements satisfying these requirements are
p-Terphenyl (PTP) with an emission curve between 280 and 350~nm, 
diphenyloxazolyl-benzene (POPOP) between 340 and 420~nm, 
diphenyloxazole (PPO) between 305 and 365~nm~\cite{Berlman},
and tetraphenyl-butadiene (TPB) between 400 and 480~nm~\cite{Berlman,burton}.

We select TPB above other waveshifting compounds
for the well matched emission spectrum to our bialkali photocathodes.
Tetraphenyl-butadiene (TPB) coatings 
have been studied in~\cite{burton, Grande:1983lr, Davies:1996p765,
McKinsey:1997p509, McKinsey:2004p1356} and successfully used in previous
experiments (see e.g. ~\cite{Amerio:2004p2123, WarpCollaboration:2008p404}).
They are particularly well suited for the detection of DUV light due to 
the large Stokes shift of TPB~\cite{Berlman,burton, Davies:1996p765, Flournoy:1994p1322}.
The fluorescence
decay time is about 1.68~ns~\cite{Flournoy:1994p1322} and,
since no phonons are involved, the recombination process does not
slow down significantly at cryogenic temperatures.
TPB coatings can be made
durable with good adherence to the substrate and 
high resistance to mechanical abrasion. The coatings are generally not 
soluble in water but can be removed when necessary 
by using toluene, chloroform ($CHCl_3$) or other organic solvents.  
TPB coatings have been exposed to high vacuum conditions
for very long periods of time and
have shown no evidence of 
significant change in the detector sensitivities.

\subsection{Reflector type and outgassing measurements}

The presence of the strong electric field at the position of the light reflector (Figure~\ref{fig:intro}(left))
demands non conductive material as the substrate for the wavelength shifter.
Our research focussed on the two materials 
ESR (Vikuiti\texttrademark\ Enhanced Specular Reflector foil) from 
the company 3M\texttrademark\ and Tetratex\textregistered\ (TTX) from the 
company Donaldson Membranes.
3M\texttrademark\  foil is a multilayer specular reflecting polymer film and as such
likely to be of high radio-purity. Its appearance is that of a polished metal 
although the material is non conducting. It has a 
specular reflection coefficient of practically 100\% 
in a large region of the optical spectrum.
TTX is an aligned polytetrafluoroethylene (PTFE) fibrous cloth and is
nearly a 100\% diffuse Lambertian reflector. 
We employ TTX in the thickness of 254\,$\mu$m which is often 
used for wrapping crystals.

%===========TTX photo===========================
\begin{figure}[htb]
\begin{center}
\begin{tabular}{c}
	\includegraphics[width=.6\textwidth]{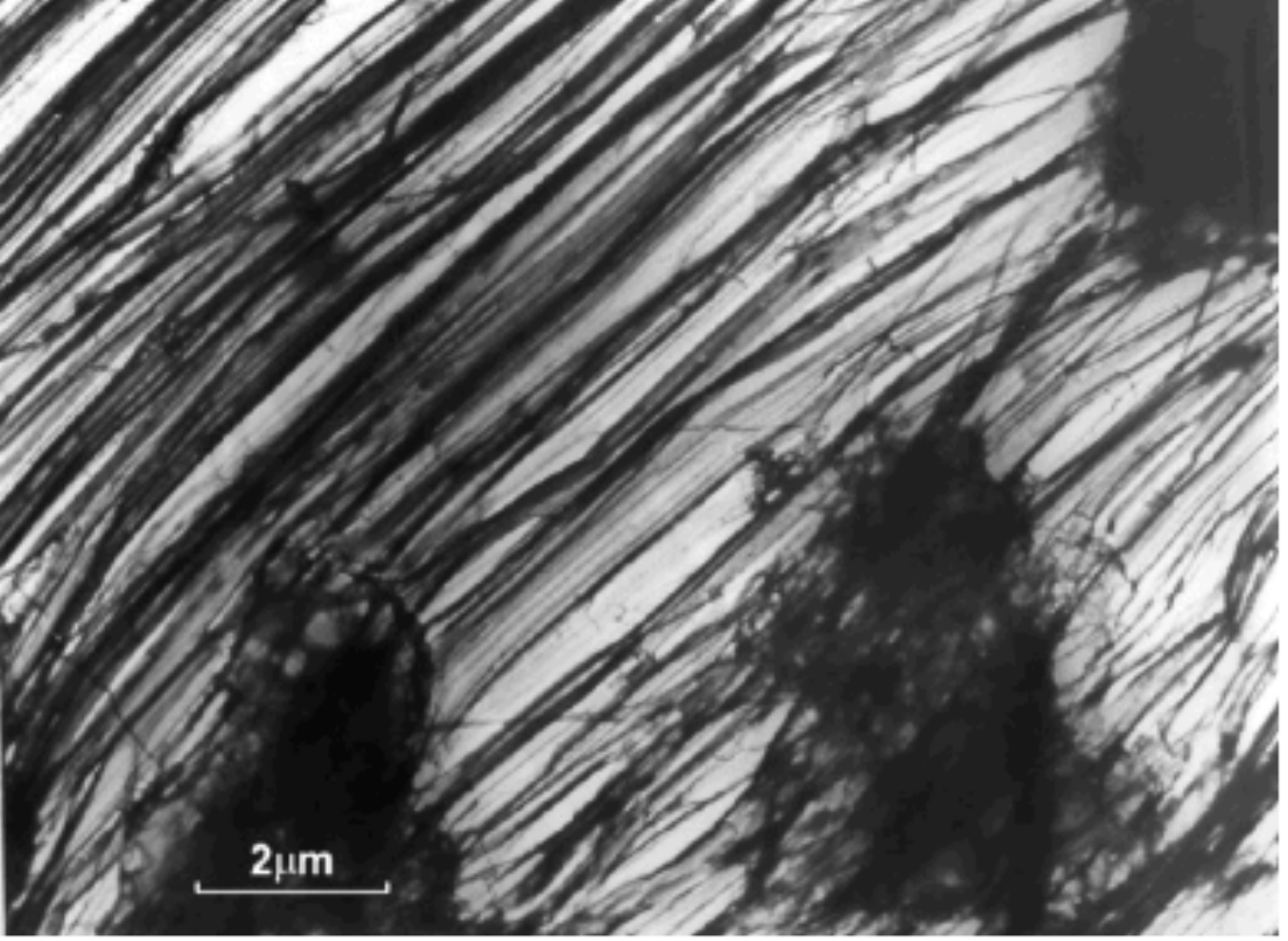}
\end{tabular}
\end{center}

\caption{\sl Photograph of a Tetratex\textregistered\ sample taken with a scanning electron microscope.}

\label{tetratex}
\end{figure}
%%--------------
A photograph of TTX cloth taken with a scanning electron 
microscope is shown in Figure~\ref{tetratex}. Because of the manufacturing method, which 
relies on extrusion of polymer chains within an oil based emulsion, doubts 
have been raised concerning both the purity of TTX and its outgassing rate. 
The outgassing properties were checked with a dedicated sample of 48\,g of 
254\,$\mu$m TTX (equivalent to 1/20 of the total amount required to line the ArDM 
experiment). The sample was placed within a large chamber which was evacuated.
A series of measurements confirmed the good outgassing properties of TTX.
The outgassing products were analysed using a mass spectrometer\footnote{MKS Spectra Microvision Plus.} and found to predominantly 
contain nitrogen and water.
In order to investigate the radio-purity of TTX, samples were sent to 
Harwell Scientifics for analysis by inductively coupled plasma mass spectroscopy 
(ICP-MS). The detection limits of the instrument are 0.3, 0.4 and 
500 parts per billion (ppb) for U, Th and K respectively. 
TTX radiopurity was found to be 1.0$\pm$0.3, $<$ 0.4 and $<$ 500 ppb for U, Th and K concentrations
respectively.
Table~\ref{radiopurity} compares the results from the 
analysis with other data taken using the ICP-MS analysis of standard target 
components. 

%===========TTX radiopurity compared table ==============
\begin{table}
\caption{\sl TTX radio-purity compared with the radio-purity of standard target components.
The reference impurity levels are based on the UK Dark Matter Collaboration internal 
measurements~\cite{ukdmc}.}
\begin{center}
\vspace{4mm}
{\footnotesize \begin{tabular}{lccc}

\toprule
Sample & U conc. (ppb) & Th conc. (ppb) & K conc. (ppm) \\
\midrule
Tetratex\textregistered\ (TTX) this paper &  1.0 $\pm$ 0.3 & $<$ 0.4 & $< $ 0.5\\
\hline
Polyethylene & $<$ 0.3 & $< $ 0.4 & $<$ 0.5 \\
Copper & 1.0 & 1.0 & 0.5 \\
Steel & 2.8 & 1.5 & $<$ 0.5 \\
Quartz & 4.0 & 4.0 & 21 \\
Borosilicate glass & 30 & 30 & 120 \\
Molecular sieve & 34 & 14 & 61 \\
Aluminum oxide & 400 & 16 & 19\\
Ceramic capacitor & 450 & 400 & 320 \\
\bottomrule
\end{tabular}}
\end{center}
\label{radiopurity}
\end{table}
%-------------------------------------------------------------------------

\subsection{Coating techniques on reflectors and PMT windows}

\begin{figure}[t]
\centering
\includegraphics[width=0.9\textwidth]{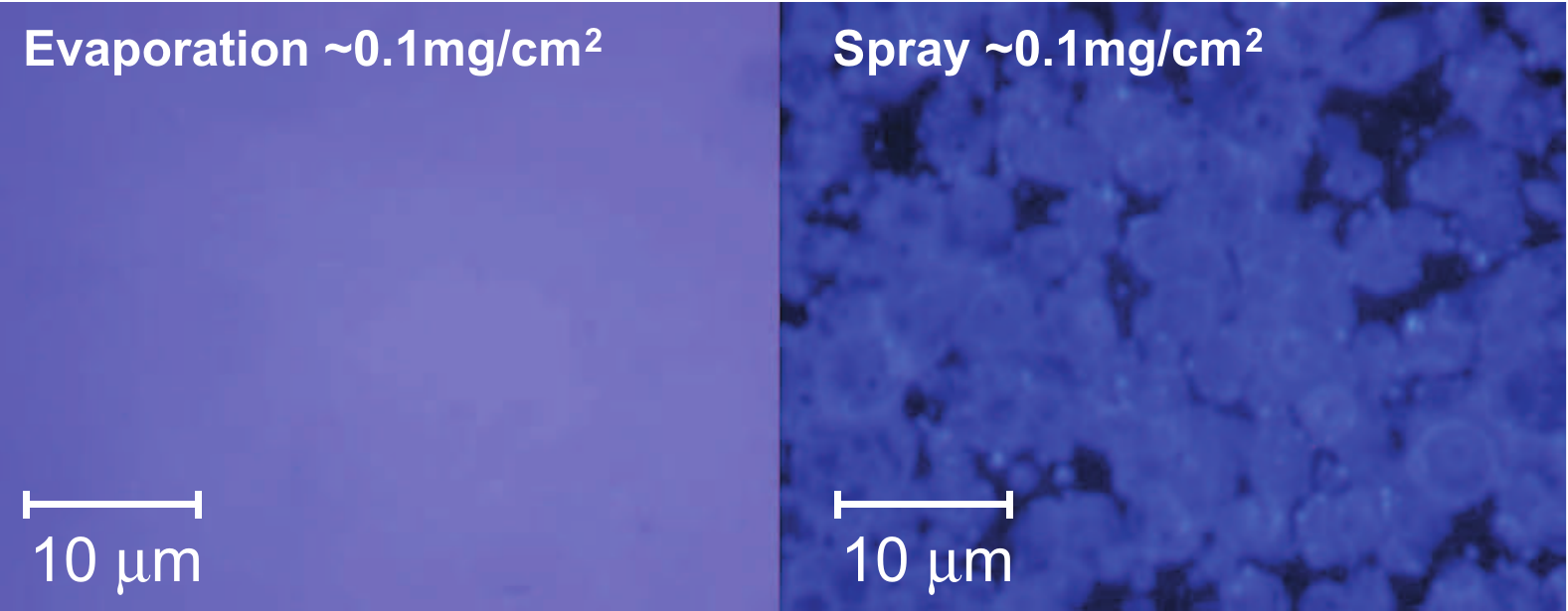}
\caption{\sl Surface structure of evaporated (left) and sprayed (right) TPB on 3M \texttrademark\ foil
under the microscope, illuminated by 250~nm light. }
\label{fig:surface}
\end{figure}

TPB powder can be applied to a reflector or PMT window by 
vacuum evaporation, spraying, or by dissolving in a polymer 
matrix~\cite{Davies:1996p765, McKinsey:1997p509, Bolozdynya:2008p1813,lorenz}. 
For comparison Figure\,\ref{fig:surface} shows microscope photographs 
of evaporated and sprayed 3M\texttrademark\  foil samples 
illuminated with UV light. The granular structure of the air brush method 
due to the formation of small crystals can easily be observed, while the coating
achieved by evaporation of TPB is very uniform.

The 3M\texttrademark\ foil samples were prepared
by evaporation in a modified 
glass exsiccator where two electrical feedthroughs were added.
%%----------------------
%%--------------------
Because of the low melting point of TPB (207\,\degr C) and its chemical 
inertness,  the requirement on the vacuum is modest. Typically 
a value of $10^{-4}$\,mbar for the residual pressure was reached.
A 15\,mm diameter Al$_2$O$_3$ crucible is employed, being able to hold up to 1.5\,g 
of TPB powder. Heating is provided by a tungsten filament wound around the 
entire height of the crucible for even heat transport. Heating power was 
set between 3--8~W
for evaporation cycles from about 30\,min to 5~hours.

For the rest of the measurements the vacuum evaporation of the samples  
was performed in a commercial Edwards model E308 evaporation chamber. 
Here up to  3\,g of TPB powder was heated by applying 24\,A current 
to the molybdenum sample holder. The reflector/PMT window was placed above the TPB 
powder at a fixed distance and the coating thickness was controlled by varying 
this distance and the weight of the powder. 

Sprayed coatings were prepared by dissolving TPB in toluene in a ratio of 1 to 
40. This solution was then airbrushed onto the substrate using 1.2\,bar argon 
gas. 

The polymer matrix coatings on PMT windows (compare also\,\cite{lorenz}) were 
prepared using long chain paraloid 
or polystyrene plastic fragments dissolved in toluene. TPB was
added and dissolved isotropically. A known amount of obtained liquid was then syringed 
onto the substrate. 
The TPB concentration within the solution was varied, as was the amount of liquid 
applied to the substrate.
The solution was left for three hours to allow the toluene to evaporate, 
forming clear TPB impregnated plastic.

%%%%%%%%%%%%%%%%%%%%%%%%%%%%%%%%%%%%%%%%%%%%%%%%%%%%%%%%%%%%

\section{Results}
\label{chap:measure}

\subsection{Method}
Preliminary investigations with gaseous argon and
\alp\ particle excitation at normal temperature and pressure (NTP) were described 
in an earlier work~\cite{Amsler:2008p1320}.  
A typical argon scintillation pulse at NTP can be described by a sum of two exponentials with the 
time constants $\tau_{1}$ and $\tau_{2}$ for fast and slow components, respectively.
The slow scintillation component can be 
used to measure the VUV light yield at 128\,nm, however,
its measured light yield and the measured decay time of 
the slow scintillation component $\tau_{2}$ strongly depends on gas purity.
This dependence is attributed to impurities destroying the long-lived triplet argon
excimer state.
Therefore $\tau_{2}$ can be used as a measure of argon purity. In order to correct for this effect
and to compare various measurements, results are
plotted versus the measured value of $\tau_{2}$. This method, described in~\cite{Amsler:2008p1320},
allows to determine the individual light yield by extrapolation to 
$\tau_{slow}$=3.2\mus, corresponding to the maximum slow decay time observed in pure argon at 1~bar~\cite{Keto:1974p593},
disentangling the effect of impurities in the argon gas.

\subsection{Reflection coefficient measurements of TPB coated reflectors }
\label{chap:spectrophotometer}

\begin{figure}[htb]
\begin{center}
\begin{tabular}{c}
	\includegraphics[width=.7\textwidth]{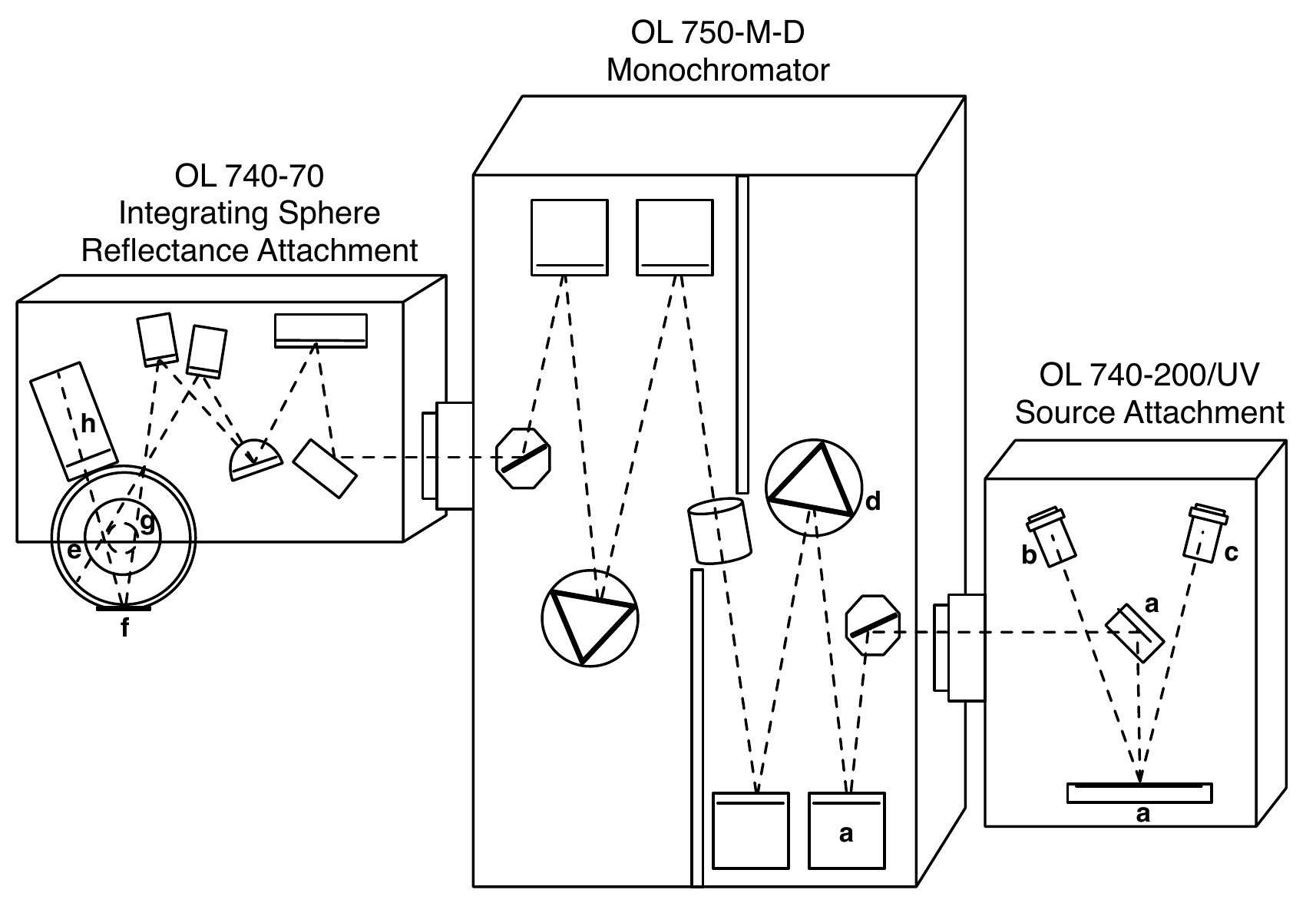}
\end{tabular}
\end{center}

\caption{\sl Schematic of the Optronics OL750 spectroradiometer.
a: selecting mirror, b: quartz halogen light source, c: deuterium light source,
d: diffraction grating mount, e:~integration sphere, f: sample, g: silicon photo diode detector,
h: light trap. Figure redrawn from the manual. }

\label{spectro}
\end{figure}

An Optronics OL750 spectroradiometer with an OL740-70 integrating sphere~\cite{optronics}
was used 
to measure reflectance and global fluorescence from 200\,nm to 700\,nm for 
a range of samples with TPB deposited by spray, evaporation, and polymerisation over 
a variety of thicknesses and concentrations, and on a range of substrates. A 
schematic of the spectroradiometer can be seen in Figure~\ref{spectro}. 
%%===========Spectroradiometer===========================
%%%------------------------

The light sources used 
were deuterium (200-400\,nm) and quartz halogen (330-700\,nm) which were 
calibrated with a mercury lamp. An automatic motorized monochromator  
allowed selecting a specific wavelength to within 1~nm. 
This light 
was directed through a small opening to be incident on a 2~cm diameter 
target sample within an integrating sphere whose internal surfaces were coated 
with PTFE. Light reflecting from the sample is then integrated within the sphere 
prior to being collected by a silicon photo diode detector.
The absolute value 
of the reflection coefficient was measured by comparing light collected from 
a sample to that produced by a calibrated NIST registered pressed PTFE 
sample\footnote{A calibrated sample obtained from the National Institute of Standards \& Technology.}. 
For diffuse reflectance measurements, the sample was positioned at 
an angle of 15~degrees to the incident light, and a light trap positioned at a 
port on the integrating sphere such that all specularly reflected light would 
be trapped and could not contribute to the integrated signal. 

%%------------
\begin{figure}[htb]
\begin{center}
\begin{tabular}{c}
	\includegraphics[width=.7\textwidth]{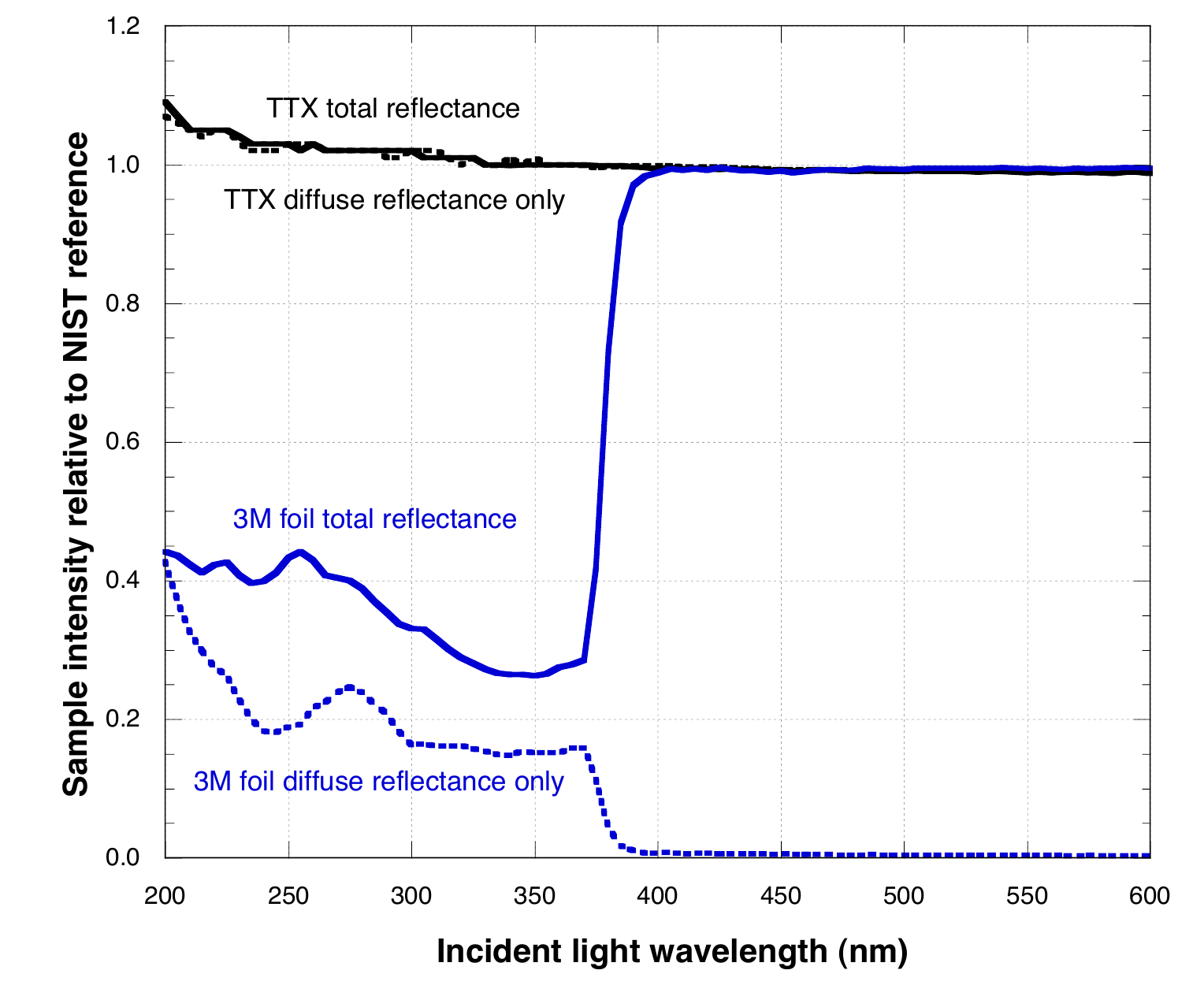}
\end{tabular}
\end{center}
\caption{\sl Comparison of total and diffuse reflection intensity for TTX and 3M \texttrademark\ foil substrates
relative to the NIST reference sample.}
\label{refl_ttx_3M}
\end{figure}

%%------------
%Figures~\ref{refl_ttx_3M} and~\ref{refl_3M}
%show the results of the analysis.

Figure~\ref{refl_ttx_3M} shows the reflectance of uncoated 3M \texttrademark\ foil and TTX cloth samples.
The wavelength on the horizontal axis refers to the incident light wavelength, selected by
the monochromator. 
The reflection coefficient cannot be measured in VUV  (i.e. below 200~nm) with this type of 
spectroradiometer, because the radiation is absorbed by air. 
In addition, below 400~nm, most light reflected off the coated samples is shifted to a mean 
wavelength of 430~nm, well within the measurable region of the equipment.
Relative reflection coef\-ficients 
greater than one indicate here fluorescence. 
%The errors on the data have not been 
%shown to preserve clarity but are estimated to be $\pm1.4$\,\%. 

As  expected, light reflected from uncoated TTX is 100\% diffuse, and light from 
3M\texttrademark\ foil above 390\,nm is 100\% specular. Interestingly, below 370\,nm light from the
3M\texttrademark\ foil is over 50\% diffuse and this rises to 100\% of its total reflectance at 200\,nm. 

\begin{figure}[t]
\begin{center}
\begin{tabular}{c}
	\includegraphics[height=0.4\textwidth]{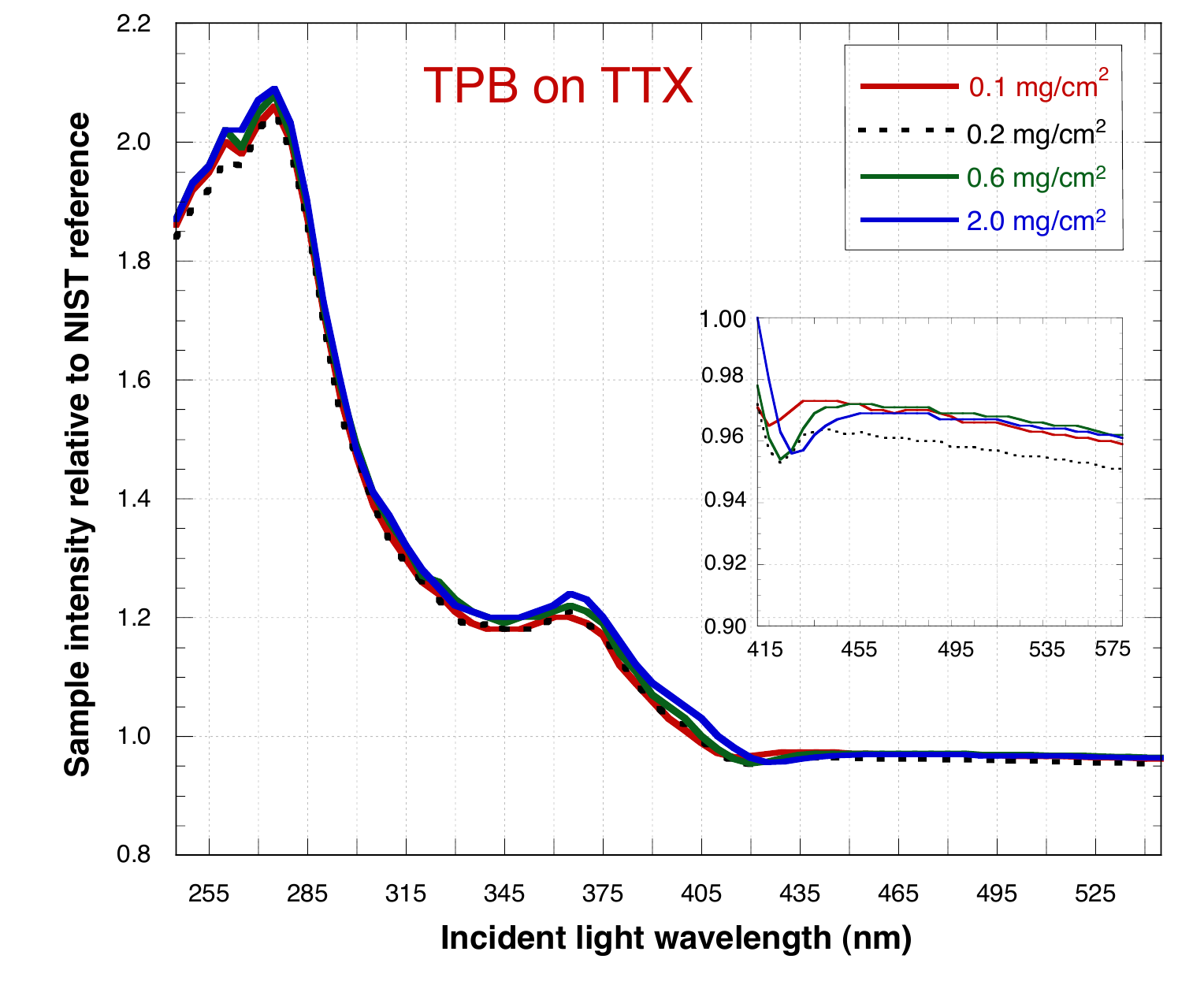}
	\includegraphics[height=0.4\textwidth]{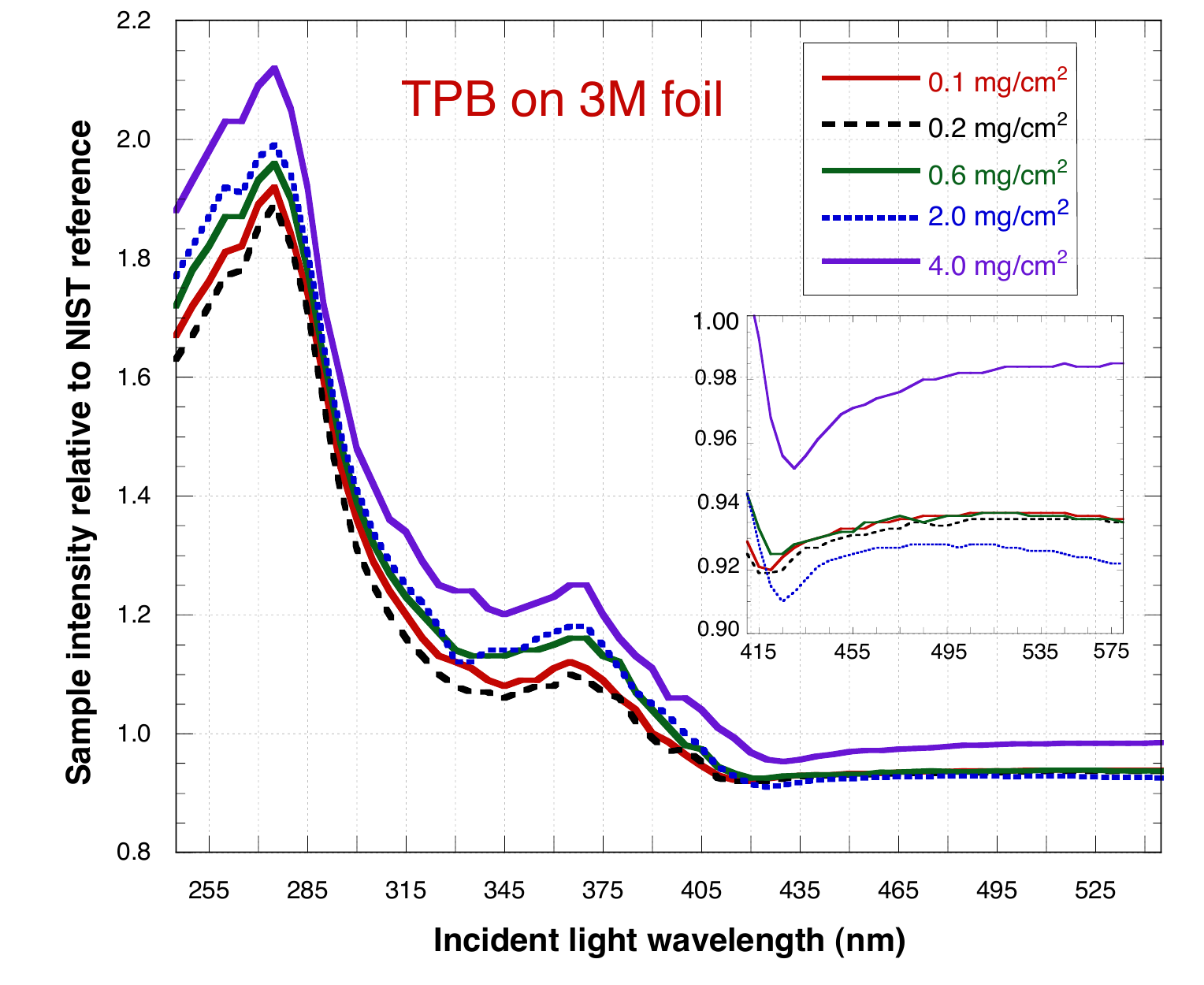}
\end{tabular}
\end{center}

\caption{\sl Total reflectance of (left) single thickness TTX cloth; (right) TPB evaporated onto 3M\texttrademark\ foil.
The insets show an expanded scale in the wavelength range 415 to 575~nm.}
\label{refl_3M}
\end{figure}

Figure~\ref{refl_3M} presents the results for different coatings. 
%The reflection coefficient for 3M\texttrademark\ foil decreases approximately by 5\% as the TPB coating 
%thickness drops from 4.0~$\mgcs$ to 0.1~$\mgcs$ at 450~nm. 
The reflectance coefficient
for TPB coated 3M \texttrademark\ foil is  about 93\% at 430~nm and varies
slightly between the thickest (4.0~mg/cm$^{2}$) and the thinnest
(0.1\mgcs ) samples.
On the other hand,  evaporated TTX 
samples have a reflection coefficient close to  97\% at 430~nm and this
property is less sensitive to the TPB thickness.

%%---------
%%------

Diffuse and total reflectances for  3M \texttrademark\ foil TPB coated 
samples were also studied.
The maximum specular component of the 
thinnest sample on 3M \texttrademark\ foil is $<$~3\% and the thickest is $<$~1\% at 430~nm. 
This indicates that irrespective of the TPB coating thickness, reflected light 
within any target will be almost completely diffuse. This result also implies 
that the 3M \texttrademark\ foil is contributing little to the reflectivity of the sample for 
the thickest coatings (which on 3M \texttrademark\ foil have consistently produced more light)
for wavelengths between 345 and 555~nm. 
These measurements favor the TTX fabric as the base of our reflectors.

\subsection{Conversion efficiency of a thin TPB layer on 3M \texttrademark\ foil}
\label{chap:convTPB}

In a first test we investigated the wave shifting efficiency of thin TPB layers on 3M\texttrademark\   foil
with respect to their thicknesses~\cite{hugo}. TPB was applied onto the samples by vacuum evaporation
in the above described exsiccator. A set of 13 disks of 3M\texttrademark\  foils (diameter 70 mm) was prepared 
with TPB layers ranging from 0.011 to 2.8mg/cm$^2$, while a 14th disk was left untouched. The amount 
of TPB deposited on the disks was determined by weighing the disks before and after evaporation.
During all evaporations the foils were held at constant distance to the crucible. The amount
of TPB loaded in the evaporator crucible was found to be in nice linear relation to the amount
of TPB on the disk after evaporation. The increase in brightness of thicker TPB
layers becomes apparent under illumination with a UV lamp (300nm). The response to
128nm light was then determined using scintillation light from argon gas. 

%%-------------------
\begin{figure}[htb]
\centering
\includegraphics[width=100mm]{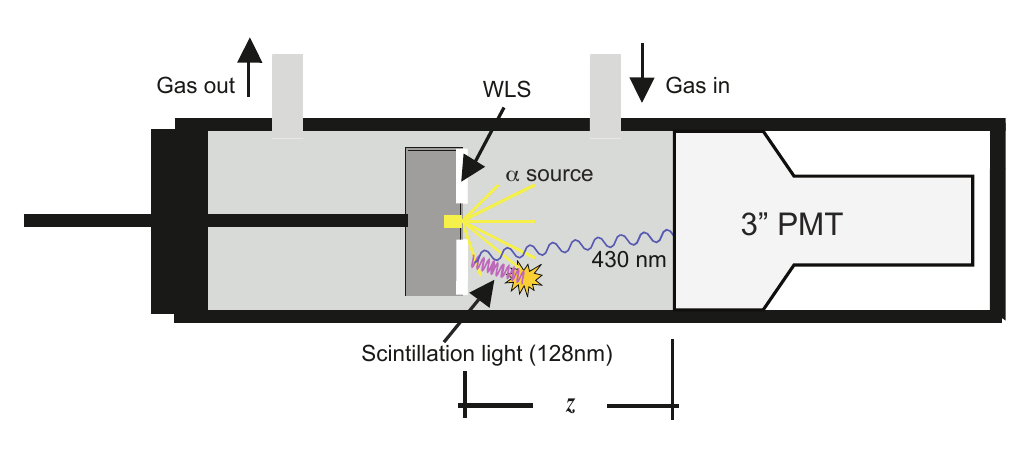}
\caption{\sl Setup to determine the conversion efficiency of TPB coated 3M \texttrademark\ foil disks.}
\label{fig:hugosetup}
\end{figure}
%%----------------------

\begin{figure}[htb]
\centering
\includegraphics[width=0.95\textwidth]{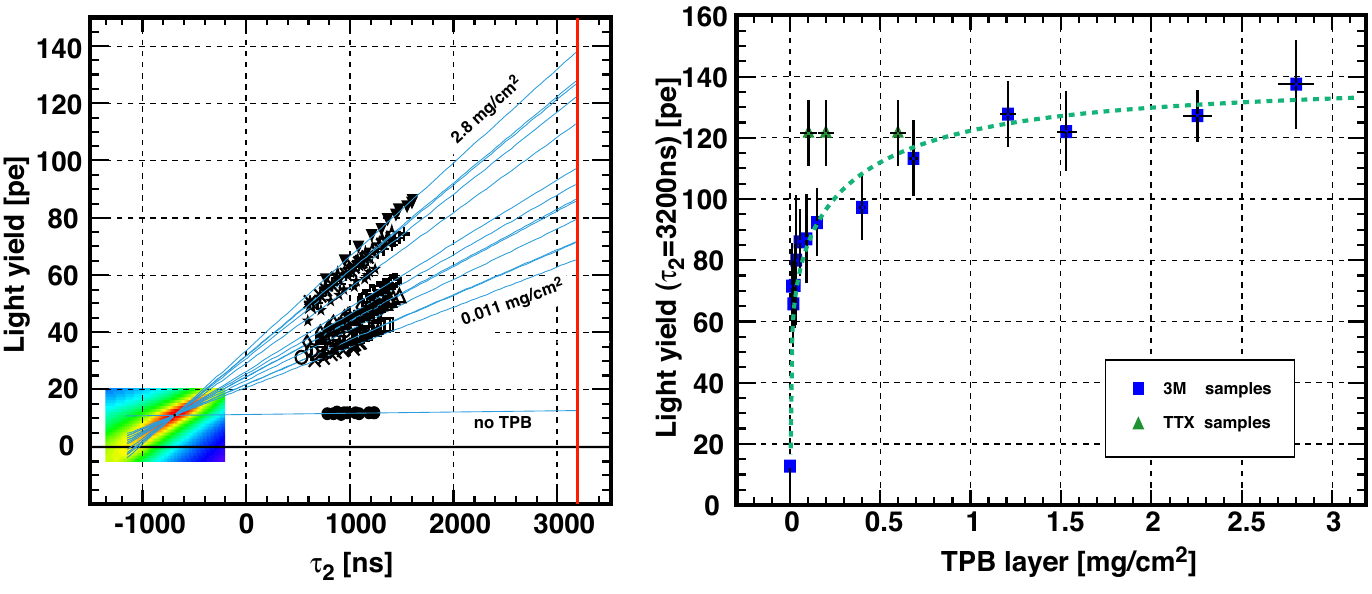}
\caption{\sl (left) Light yield in photoelectrons vs lifetime of the slow scintillation 
component for various thicknesses of TPB coatings on 3M \texttrademark\ foil; 
(right) Conversion efficiency (expressed as absolute yield in photoelectrons for an effective slow component
of\, $3.2~\mu$s) for the various thicknesses on 3M\texttrademark\ foil compared
to a few measurements of TTX foils.}
\label{fig:plots}
\end{figure}

For this purpose the
foils were mounted onto a movable holder opposite to an uncoated bialkali PMT ($z$\, =\, 6.7 cm)
in a black lined tube (Figure \,\ref{fig:hugosetup}) which was connected to a high purity (99.9999\%) argon gas 
system. Without prior pumping the gas flow improves gradually the argon purity in the chamber
while the scintillation light from $^{241}$Am $\alpha$-particles is recorded continuously\footnote{The total deposition of $\alpha$ energy occurs within 4\,cm in 
1\,bar of gaseous argon.}.
The average photoelectron number from $\alpha$ signals of individual data sets was then 
plotted vs. the measured value of $\tau_{2}$, the decay time of 
the slow scintillation component (See Figure\,\ref{fig:plots}(left)).
Every line corresponds to a different disk and hence a different TPB thickness. 
Deviations of measurement points from the lines are caused by not stable starting 
conditions (mainly air humidity) and uneven increase of the argon purity among the taking 
of the different data sets. 

The lines were fitted to a common intersection point which was left floating. The colour map 
represents the  $\chi^{2}$-values of this point (red = smallest  $\chi^{2}$). 
The x and y -positions of this point reflect a contribution of non 128nm (but UV) light from fast light 
emission in NTP argon (roughly 17\% at maximum gas purity [8]) and background of residual VUV
sensitivity, respectively. The extrapolated (purity normalised) light yields (i.e. intersections 
of the lines with the value of 3.2$\mu$s for $\tau_2$) are plotted vs. the amount of TPB on the disks in 
Figure\,\ref{fig:plots}(right). The result (errors from the line fits) is consistent with a saturation of 
VUV conversion efficiency above 1 \mgcs (the dashed line is meant to guide the eye). 
A very similar result is achieved by not constraining the lines to a 
common intersection point but produces larger fluctuations. 

%%----------------------
%%--------------------------- 

The measurements were repeated on three TTX samples (triangles in Figure~\ref{fig:plots}(right)) 
showing a generally good performance and again a surprisingly weak dependance on
the TPB layer thickness (consistent with the measurements
of Section~\ref{chap:spectrophotometer}).

\subsection{Global efficiency of wavelength shifting and reflection}
\label{chap:globaleff}

\begin{figure}[htbp]
\begin{center}
\begin{tabular}{c}
	\includegraphics[width=.7\textwidth]{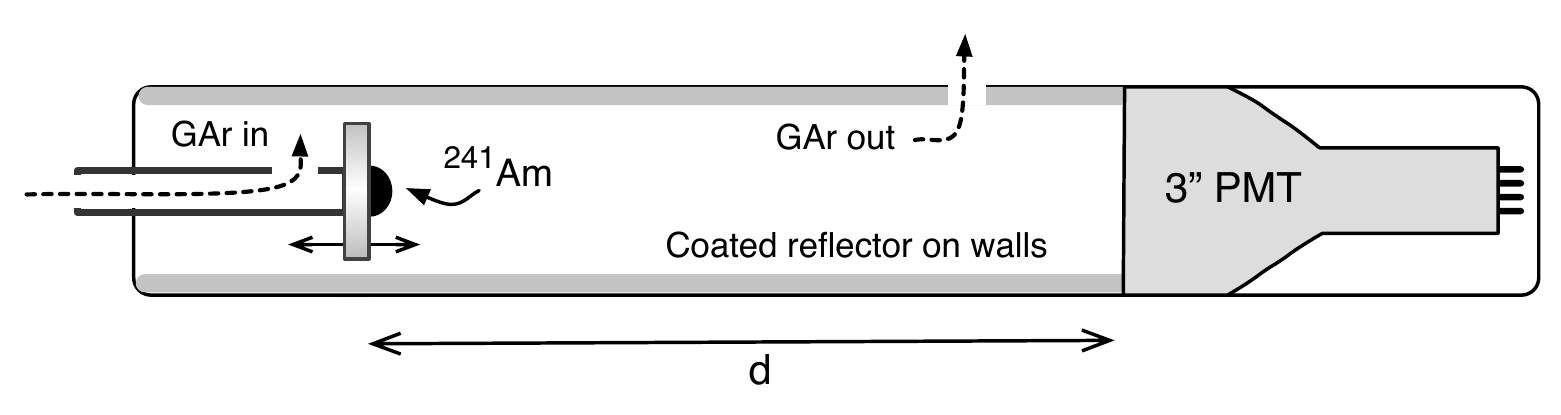}
\end{tabular}
\end{center}

\caption{\sl Schematic illustration of the argon gas apparatus 
used to determine the wavelength shifting and reflection efficiency 
as a function of distance $d$ (ranging between 300~mm and 920~mm).}

\label{w_apparatus}
\end{figure}

A schematic illustration of the argon gas apparatus which was used
is shown in Figure~\ref{w_apparatus}.
This apparatus consisted of a sealed polyvinyl chloride (PVC) tube containing a 
3'' uncoated ETL PMT type 9302KB. An $\alpha$-source located at the centre 
of a TPB coated reflector disk was placed within the tube at a variable distance 
from the PMT. Samples of either 3M \texttrademark\ foil or TTX cloth coated with TPB 
were placed around the interior walls of the tube. A gas delivery tube was inserted 
into the PVC chamber and 99.9999\% pure argon gas at 1\,bar flowed through
the apparatus. The argon flow rate was used to control the argon purity. 
Measurements were taken for various TPB thicknesses between 0.2\mgcs\ 
and 4.0\mgcs , which were deposited both via evaporation and spraying. 
Additionally, the distance between the $\alpha$-source and the PMT was altered 
in order to investigate the effect of both the attenuation of light following 
multiple reflections and the reduction in direct VUV light incident on the 
PMT. The total number of photoelectrons collected by the PMT 
was analyzed for various distances $d$ in the same way as
described in Section~\ref{chap:convTPB}.

%% -----------------------------------
%% Argon gas apparatus figure
%% -----------------------------------
%% -------------------------

%%%---------
%\begin{figure}[htb]
%\begin{center}
%\begin{tabular}{c}
%	\includegraphics[width=0.9\textwidth]{figures/phot_sep3200b}
%\end{tabular}
%\end{center}

%\caption{\sl Total number of photoelectrons extrapolated to slow decay times of  $3.2~\mu$s plotted against separation $d$ from the 
%$\alpha$-source to PMT, for TPB coated reflector walled tube.}

%\label{3200nspurity_separation}
%\end{figure}
%%%----------

Evaporated coatings on 3M \texttrademark\ foils 
consistently underperformed TTX cloth irrespective of coating thickness. 
Thicker coatings on 3M \texttrademark\ foil yielded 
higher light collection whereas light collection from TPB evaporated on TTX 
substrates was found to be almost independent of thickness. The 0.2~\mgcs\ TPB 
on TTX yielded within errors an identical result as for the 1.0~\mgcs\ coating.

A visible difference between 0.2~\mgcs\ sprayed and evaporated TPB on 
TTX cloth implied that spraying produces areas of low coating thickness 
and large inhomogeneity, while deposition via evaporation produces
an uniform coating. 

We concluded that it is
advantageous to use evaporated TPB on TTX since smaller thicknesses of TPB 
may be used compared to 3M \texttrademark\ foil which requires between 4 and 20 times the 
thickness for comparative light collection. Thick coatings are far more brittle 
and fragile at low temperatures or when bent. This is especially true on 3M \texttrademark\ foil which,
unlike TTX, does not provide any substantial keying to the surface.

TTX cloths coated by evaporation with 1.0~\mgcs\ TPB were finally chosen for the ArDM reflector.

\subsection{WLS coating on the PMT window}
\label{chap:wlspmt}

\begin{figure}[htbp]
\begin{center}
	\includegraphics[width=.35\textwidth]{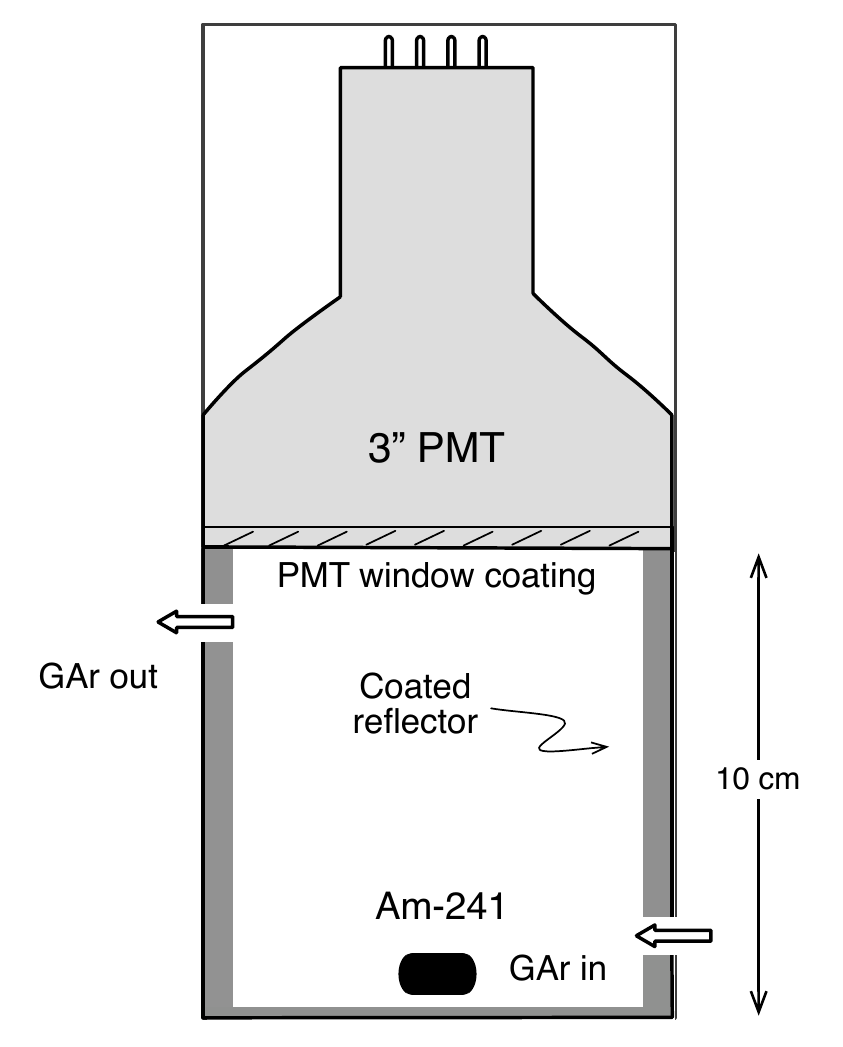}
\end{center}
\caption{\sl Schematic illustration of the argon gas apparatus used 
to determine the wavelength shifting efficiency of direct light incident
on a TPB coated PMT.}
\label{GAr_ap2}
\end{figure}

Another argon gas apparatus for direct light measurements was constructed
with a similar aspect ratio as the
full scale ArDM detector (Figure~\ref{GAr_ap2}). The experiment consisted of a sealed 
PVC tube containing a coated 3'' ETL PMT type 9302KB. The PMT window was coated with 
TPB powder with thicknesses ranging from 0.02\mgcs\ to 2\mgcs\ via 
evaporation, spraying and application of a polymer matrix containing TPB. 
The sides and base of the PVC tube were covered with 3M \texttrademark\ foil reflector coated 
with 1\mgcs\ TPB powder by evaporation. 
TPB coated reflector walls were used as the ability of the window coating to shift VUV 
light is equally important as its ability to allow shifted visible light from the 
walls to penetrate. 
An $\alpha$-source was positioned 
10\,cm away from the PMT window and argon gas was flowed continually.

TPB coating thicknesses above 2\mgcs\ are not recommended since 
the transparency of the PMT window coating to visible light would be significantly reduced. 
To avoid TPB crystallisation, deposition by spraying must be slow allowing 
for the evaporation of toluene. For polymer matrix coatings, crystallisation of 
TPB could be avoided with the addition of a plasticiser, which cross-links 
paraloid/polysterene chains with TPB, thus forming a rigid lattice while the 
solvent evaporates. 
Different coating thicknesses were tested for each one of the coating 
techniques~(evaporation, spraying, paraloid and polystyrene matrix).
%%========GAr apparatus 2==========================
%%--------------
%Figure~\ref{pmtface_bestcoatings}
%presents the best results for each coating technique. 
From the different measurements, TPB coating by evaporation and
TPB embedding in a polymer matrix (e.g. paraloid) were found
to be the best within errors. The polymer matrix can be doped in a way that 
thin, crystal clear wavelength shifting coatings are produced.
In general, PMT coating improves the total light 
collection in our geometry
by about 30\% (see also Ref.~\cite{Amsler:2008p1320}). 

%%%-----------
%\begin{figure}[htbp]
%\begin{center}
%\begin{tabular}{c}
%	\includegraphics[width=.9\textwidth]{figures/bestpmtcoat}
%\end{tabular}
%\end{center}

%\caption{ Measured light yield (in photoelectrons) as a function of the measured $\tau_2$ for
%several different PMT window coatings (polysterene and 
%paraloid matrices, evaporation and spray). {\bf --- AR: plot from -500 to 3200 ns ---}}

%\label{pmtface_bestcoatings}
%\end{figure}
%%%------------------

%%%%%%%%%%%%%%%%%%%%%%%%%%%%%%%%%%%%%%%%%%%%%%%%%

\section{Application to the ArDM detector case}

\subsection{The large evaporation chamber}
\begin{figure}[htb]
\centering
\includegraphics[width=.55\textwidth]{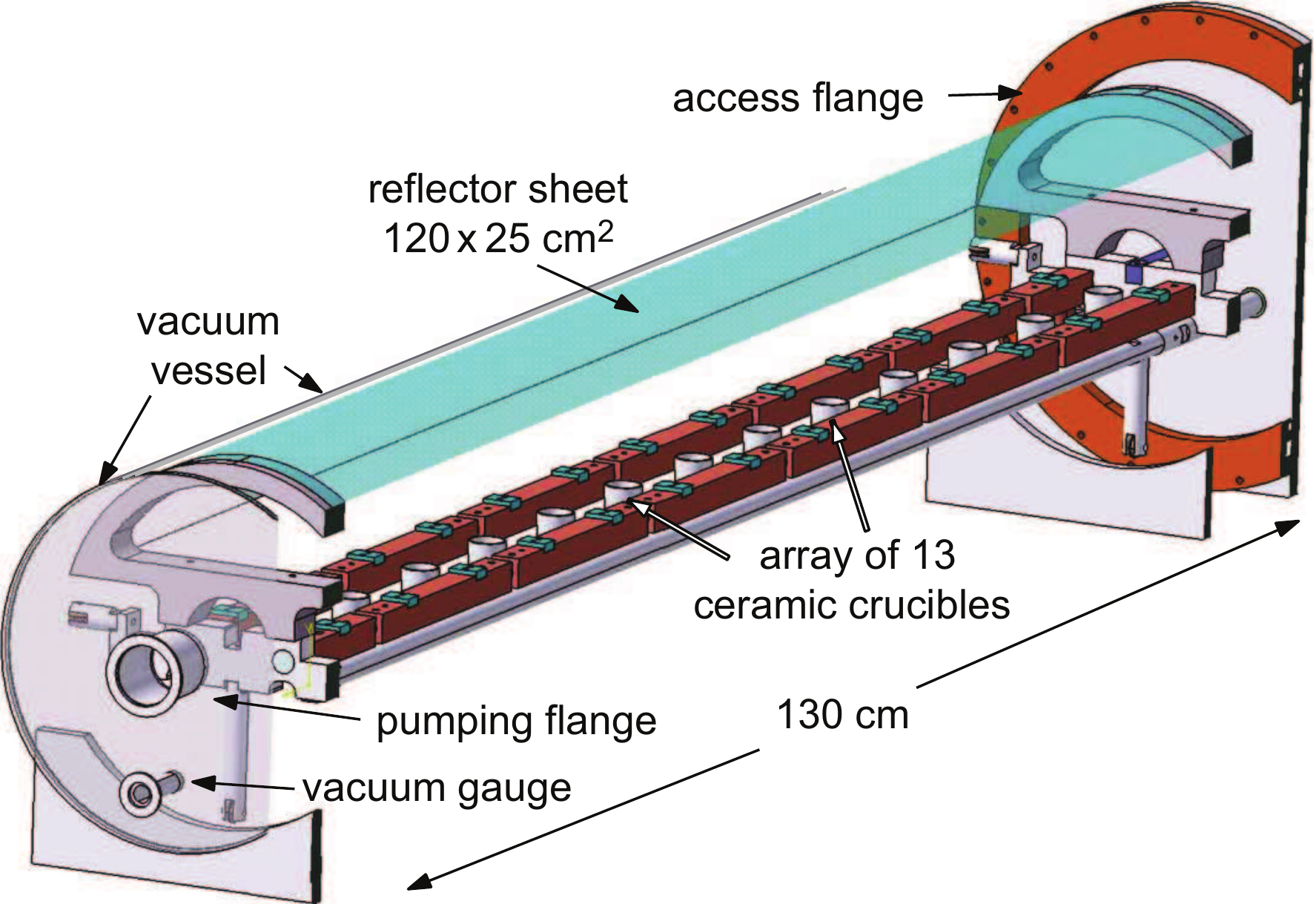}
\caption{\sl 3D sketch of the evaporation apparatus for the reflector sheets of 
the ArDM detector.}
\label{fig:largescaleEVAP}
\end{figure}

Driven by the experience gained with the different TPB deposition techniques,
described in the previous sections,
we built a stainless steel vacuum chamber large enough to house
single reflector sheets of 120$\times$25\,cm$^{2}$ for the ArDM detector.
The apparatus consists mainly of two parts, a horizontal tube with pumping 
connection on its closed end and a slide-in array of 13 crucibles mounted onto 
a Viton-sealed access flange. The crucibles are electrically connected 
in series for better uniformity and lower total supply current ($\approx$10\,A).
The reflector sheets are supported by 100\,$\mu$m wires in a crescent 
arrangement for constant distance to the crucibles (see Figure\,\ref{fig:largescaleEVAP}). 
An evaporation cycle was started by filling the crucibles with TPB powder 
and positioning the TTX reflector sheet on its support. 

%%-------------------------
%%--------------------------------

\begin{figure}[ht]\begin{center}
\includegraphics[width=0.65\textwidth]{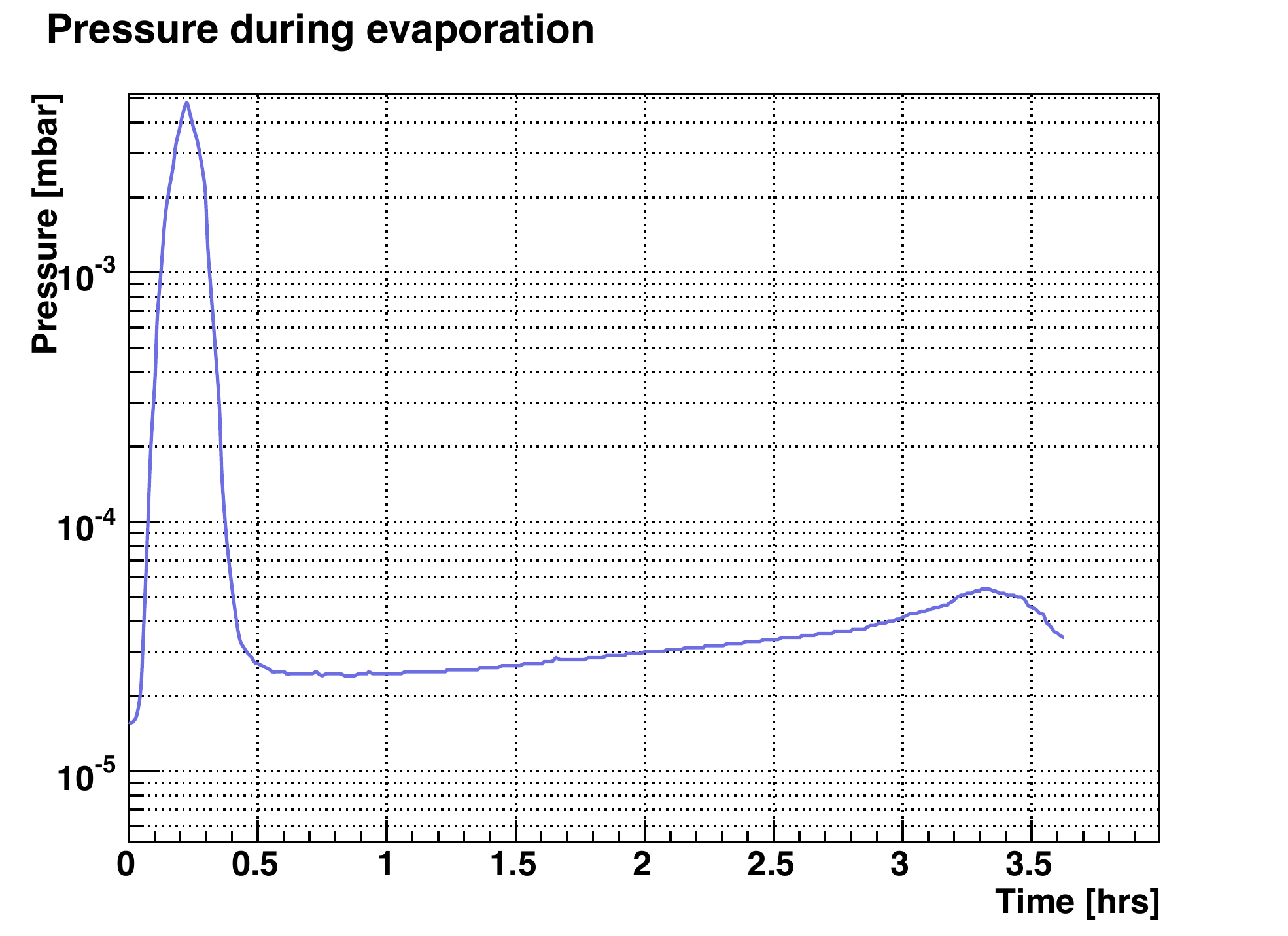}
\caption{\sl During the evaporation we constantly monitored the pressure of the evaporation vessel. The second peak shown in the graph, roughly  after $\approx 4$ hours, is interpreted as the moment in which no more TPB is remaining in the crucible.}
\label{EvapPressure}\end{center} \end{figure}

The reflector sheet surface was gently wiped by means of a grounded antistatic brush before
the apparatus was closed and pumped. After reaching a typical pressure of $10^{-5}$~mbar
the heating current was switched on and left running for about $3\sim5$ hours. The end of TPB
evaporation was signaled by a small drop in the monitored vacuum pressure, presumably
due to a small fraction of adsorbed water to the TPB powder (see Figure~\ref{EvapPressure}).

\begin{figure}[ht]\begin{center}
\includegraphics[width=0.49\textwidth]{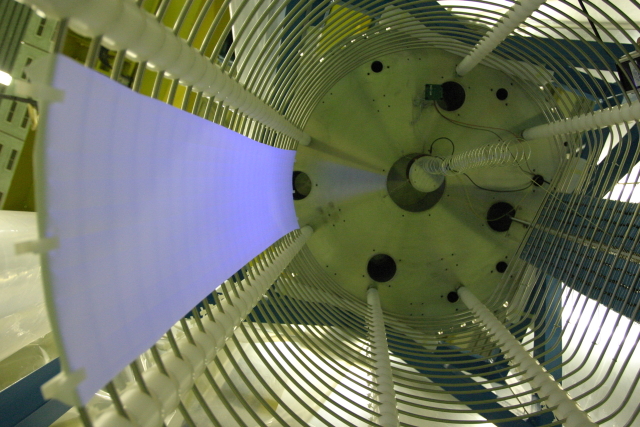}
\includegraphics[width=0.49\textwidth]{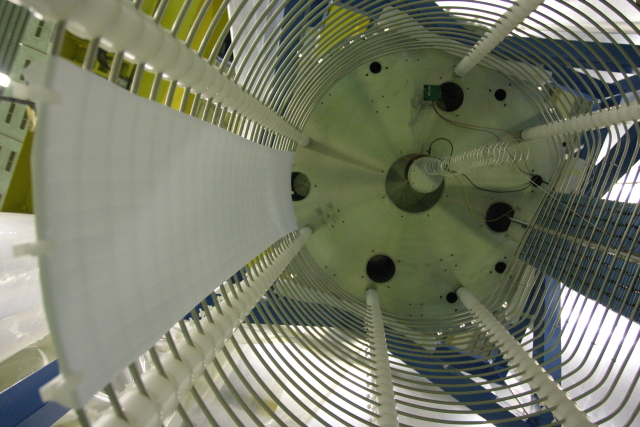}
\caption{\sl Optical inspection of one reflector sheet with a UV lamp (left) with illumination (right) without. The shifted light is readily visible
on the illuminated reflector.}
\label{fig:OpticalInspection}\end{center} \end{figure}

After this preparation the sheets were inspected optically with a UV lamp (see Figure~\ref{fig:OpticalInspection}) and stored in a clean cabinet until installation at the experiment. No degradation in
the optical performance was observed while storing the reflector sheets up to 3 months.

\subsection{Assembly of the wavelength shifting light reflectors in ArDM}
\begin{figure}[ht]\begin{center}
\includegraphics[width=0.8\textwidth]{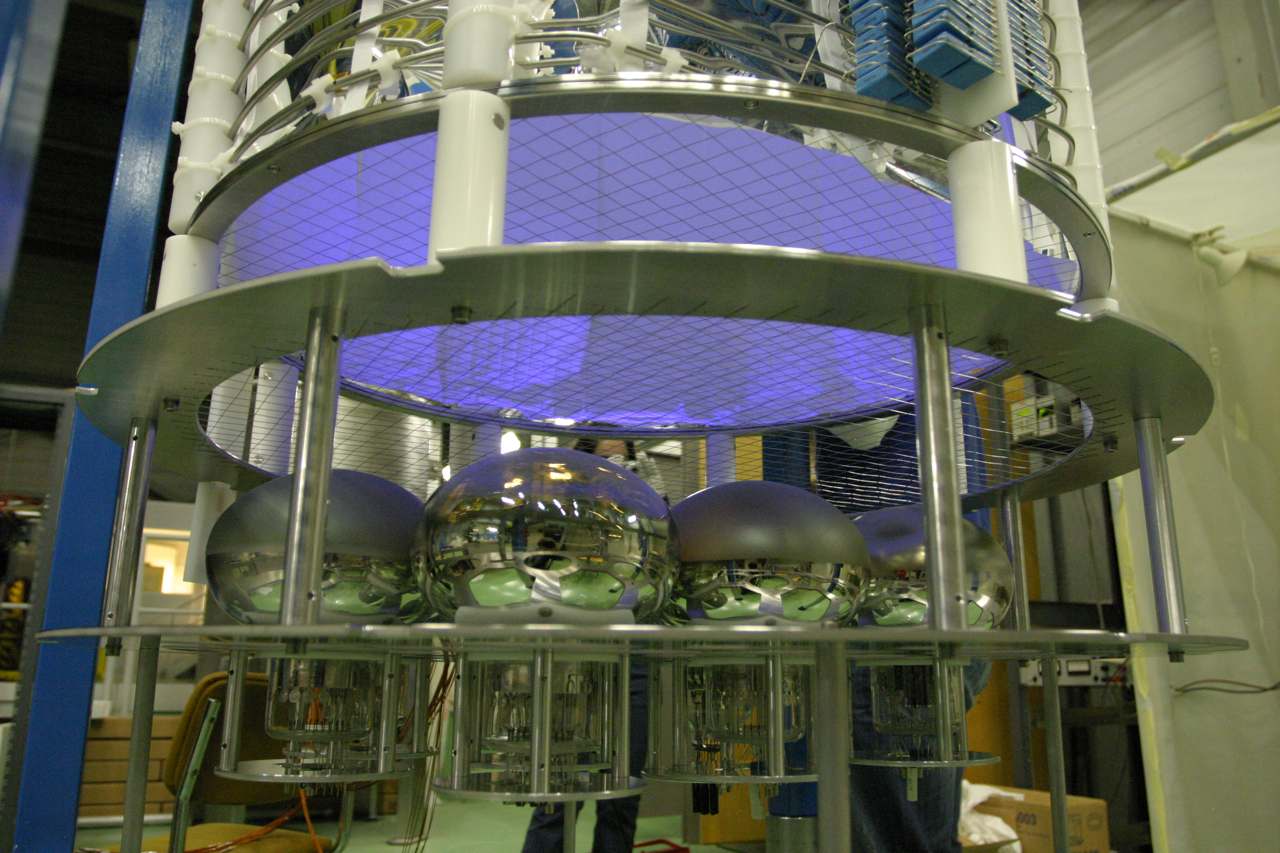}
\caption{\sl Picture of the light readout illuminated by UV light.}
\label{fig:OpticalInspection2}\end{center} \end{figure}

The wavelength shifting side reflector is made of 15 overlapping panels,  each of which  is composed of a TTX reflector installed on 3M foil substrate which acts as a substrate,  providing mechanical support and avoiding curling of the edges due to the high thermal shrinking coefficient of TTX. The panels were fixed on the first and last field shaper rings by means of teflon clips.
The reflectors were installed such that they overlap slightly on their edges and produce a uniform reflector around the detector volume. UV lamp illumination results
in a clear fluorescence (See Figure~\ref{fig:OpticalInspection2}).

The mechanical support of the photomultipliers can easily be adapted to support most of the commercially available 8" hemispherical PMTs.
Bialkali photocathodes become insulant at low temperature; for this reason this kind of photomultipliers need an additional Pt deposition on the photocathode to restore the charges at cryogenic temperature with a drawback of a reduced quantum efficiency ($\simeq 2/3$ of the original). 
For the first test in gas and liquid argon 8 out of 14 cryogenic PMT modules were installed, as shown in Figure~\ref{fig:OpticalInspection2}: 
seven cryogenic Hamamatsu R5912(-02)MOD PMTs (five with 14 dynodes and two with 10 dynodes) and one ETL 9357 KFLB. 
Both types were previously studied at cryogenic temperatures~\cite{bueno2008,Ankowski:2006xx}.
In addition, the PMT modules were tested in LAr (88K) with a blue LED ($\lambda_{pk}$ = 400 nm) light before 
coating and assembly in the experiment.

Ten PT-1000 temperature sensors, placed at different depth in the detector, allowed precise measurements of the inner temperature of the detector.
The electric field generator chain and the E-Field protection grid, which protects the PMT surfaces from the high potential of the cathode,  were installed but not used for these tests.

An $^{241}$Am $\alpha$-source (40 kBq) on a movable magnetic
actuator 
and two blue LEDs were installed in the center of the main flange. An optical fiber was glued onto one of the LEDs;  
the other end of the fiber was mounted on the side of the movable source. In this way the source and the fiber can 
be moved vertically $\pm 30$ cm from the center of the detector which allows the measurement of the light yield at different "event" positions. 
This source has a very thin metallic window which led to a reduction of the energy of the $\alpha$s from 5.486 MeV to  4.5 MeV whose mean path is  $\approx 3.5$ cm in gaseous argon at 1.1 bar.

The signals of the PMTs are digitized by independent ADC channels (1GS/sec, 12bit resolution), triggered by a fast multiplicity signal from a  programmable low threshold discriminator VME module. 
Data analysis is performed off-line, where low energy events from natural radioactivity as well as high energy cosmic ray muons can be easily discerned.

\subsection{First measurement in gaseous argon}
\begin{figure}[thb!]\begin{center} 
\includegraphics[width=0.8\textwidth]{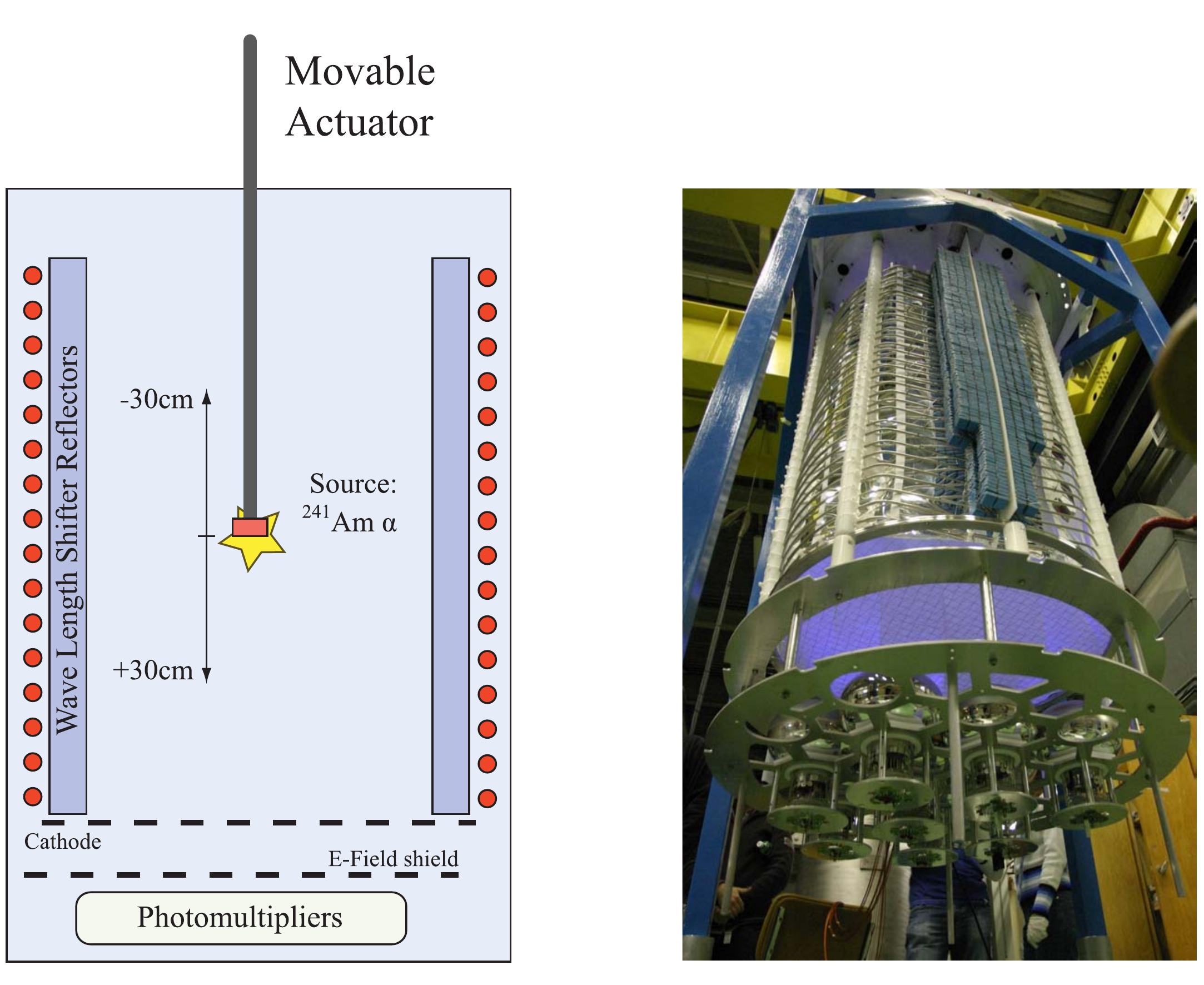}
\caption{\sl \label{setupt} (left) Schematic of the setup used in the first tests; (right) Photograph of the assembled setup with internal UV light illumination.}
\end{center}\end{figure}

The test setup is schematized in Figure~\ref{setupt} (left) while the picture of the assembled setup illuminated by UV light from inside is shown in Figure~\ref{setupt} (right).

The setup was first closed and evacuated down to $10^{-5}$ mbar without baking. The dewar was then filled with
pure gaseous argon\footnote{Argon N60 or ALPHAGAZ\texttrademark 2 ($>99.9999\%$) argon, from Carbagas AG,  ($\le 1$ ppm)} up to a pressure of 1.1 bar.

At the beginning we studied the performance of the photomultipliers: noise, dark count rate, and calibration curves
were measured. Unfortunately two of the photomultipliers could be operated only at low supply voltages (and consequently low gain) 
in gaseous argon because of sparking and could not be precisely calibrated (they will not be included in the following analysis).

\begin{figure}[ht]\begin{center}
\includegraphics[width=0.95\textwidth]{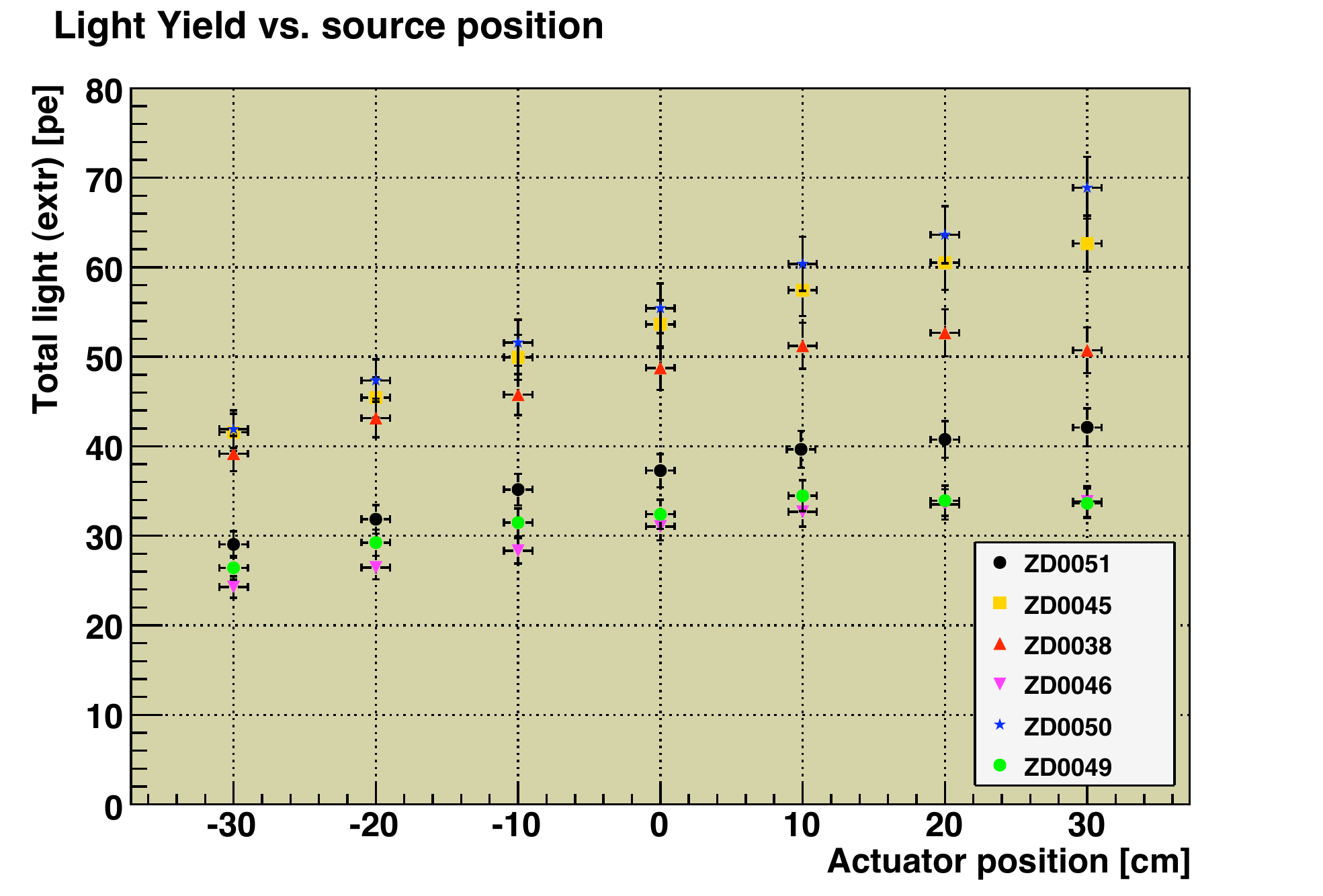}
\caption{\sl Average light detected with 6~PMT (labelled ZDxxxx) from 4.5 MeV $\alpha$ particle in GAr at 1.1bar  for different source positions extrapolated to a slow component decay time of 3200ns.}
\label{serie7}\end{center} \end{figure}

At this point we acquired several series of data with the source at different position in the detector as can be seen in Figure~\ref{serie7}. Comparison of data with Monte Carlo is ongoing.

Argon gas was then left in for 1 month. For other 2 months the setup was operated with GAr at various pressures up to 2 bar. The setup was then again evacuated down 
to $10^{-5}$~mbar, filled with pure argon and the measurement was repeated.

\begin{figure}[ht]\begin{center}
\includegraphics[width=1\textwidth]{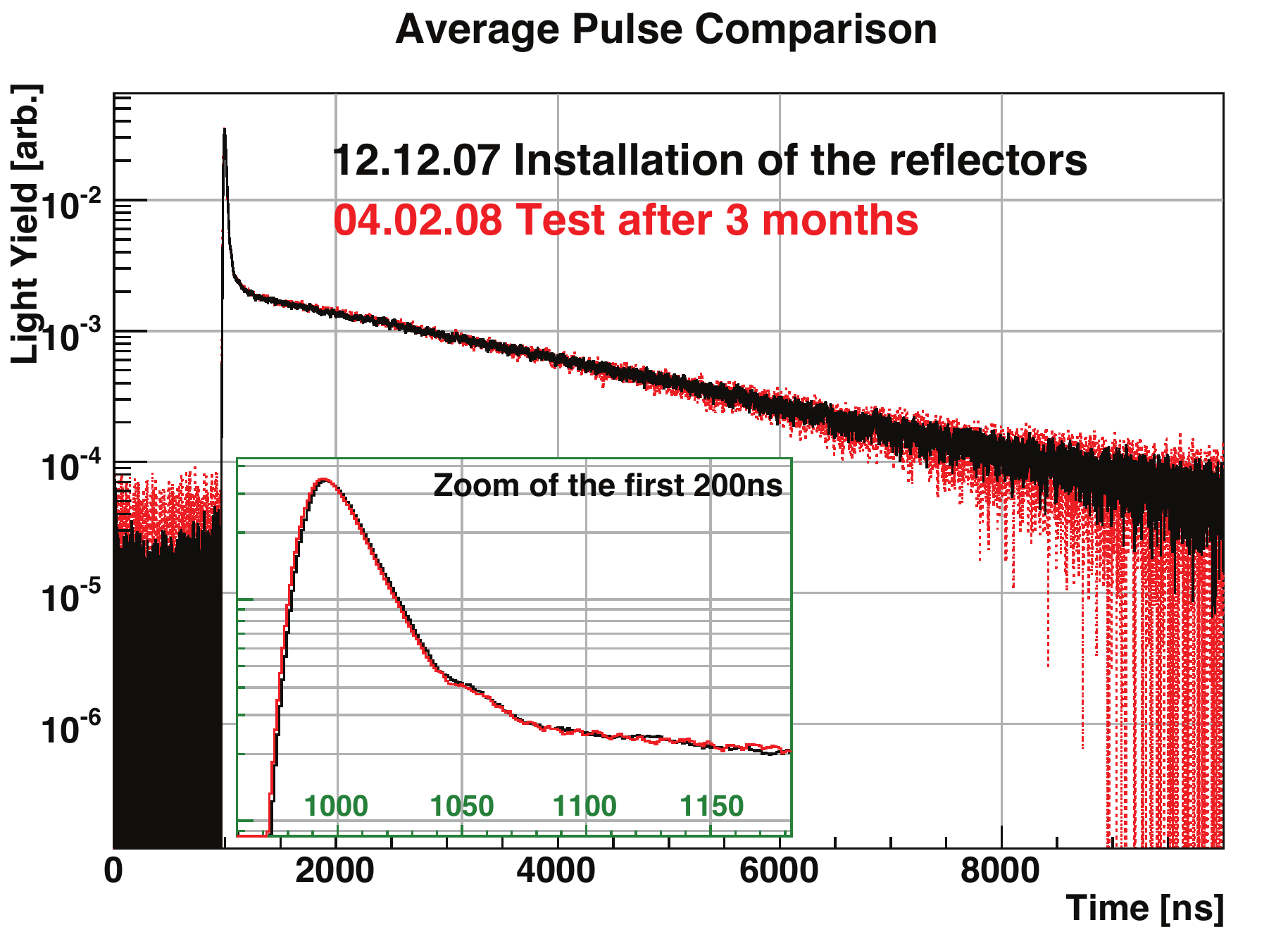}
\caption{\sl Signals from 4.5 MeV $\alpha$ particle (average pulse shape) in GAr at 1.1bar for one of the photomultipliers. The black and red lines correspond to the signal measured just after the installation of the reflectors and after 3 months of operation, respectively.}
\label{aging}\end{center} \end{figure}

We compared then the average pulse shapes of Am $\alpha$-particles of two different data sets which had the same purity condition (same decay time of the slow component) and identical source position.
The two data sets, as can be seen from Figure \ref{aging}, have the same amount of light within 1$\%$.  In conclusion there is no evidence of
change in the light yield over a time interval of 3 months.

%%%%%%%%%%%%%%%%%%%%%%%%%%%%%%%%%%%%%%%%%%%%%%%%%

\section{Conclusions}

TPB can efficiently absorb VUV radiation from argon luminescence and re-emit
at a wavelength which can be readily detected by bialkali photocathodes.
In addition the wavelength shifting process is fast and insensitive to 
cryogenic temperatures. 
The optimum TPB deposition method was found to be
vacuum evaporation which avoids crystallisation and coating inhomogeneities
typically created during spraying. TPB coated
TTX cloth was found to be the superior reflector when compared with
3M\texttrademark\  foil
due to its better light yield and 
greater tolerance to TPB layer thickness.
Based on spectroradiometer measurements TPB coated TTX was found to
have a reflectance coefficient at 430~nm close to 97\% for all coating thicknesses.
These measurements were used to define the parameters for the reflectors
of the ArDM detector. The coating and reflector combination was chosen to be
1.0~$\mgcs$ TPB deposited via evaporation on TTX cloth.
Fifteen large $120\times 25$~cm$^2$ TTX sheets
were coated and assembled in the detector. Preliminary measurements in warm 
gas were performed with an $^{241}$Am $\alpha$-source. The light yields
were measured at different time intervals, showing
no evidence of aging in the time interval of 3 months.

\acknowledgments
This project is supported by ETH Zurich and the Swiss National Science
Foundation (SNF).

% ==============================================

% ==============================================

\end{document}